\documentclass[5p, sort&compress]{elsarticle} 
\usepackage{graphicx}
\usepackage{amssymb}
\usepackage{epstopdf}
\usepackage[amssymb,thinqspace,thinspace]{SIunits}
\usepackage{amsmath}
\usepackage{booktabs}
\usepackage{rotating}
\usepackage{multirow}
\usepackage{psfrag}
\usepackage{adjustbox}
\usepackage{csquotes}
\usepackage{arydshln}
\usepackage{textgreek}
\usepackage{color}
\usepackage{ulem}
\usepackage{adjustbox}
\usepackage[super]{nth}
\usepackage{comment}
\usepackage{dblfloatfix}
\usepackage{url} 

\journal{arXiv}

\begin{document}
\begin{frontmatter}
\title{The relative contributions of TWIP and TRIP to strength in fine grained medium-Mn steels}
\author[IC]{T. W. J. Kwok}
\author[UOS]{P. Gong}
\author[IC]{R. Rose}
\author[IC]{D. Dye\corref{cor1}}\ead{david.dye@imperial.ac.uk}
\cortext[cor1]{Corresponding author}
\address[IC]{Department of Materials, Royal School of Mines, Imperial College London, Prince Consort Road, London SW7 2BP, UK}
\address[UOS]{Department of Materials Science and Engineering, The University of Sheffield, Western Bank, Sheffield, S10 2TN, United Kingdom}

\begin{abstract}

A medium Mn steel of composition Fe-4.8Mn-2.8Al-1.5Si-0.51C (wt.\%) was processed to obtain two different microstructures representing two different approaches in the hot rolling mill, resulting in equiaxed vs. mixed equiaxed and lamellar microstructures. Both were found to exhibit a simultaneous Twinning Induced Plasticity and Transformation Induced Plasticity (TWIP$+$TRIP) mechanism where deformation twins and $\alpha'$-martensite formed independently of twinning with strain. Interrupted tensile tests were conducted in order to investigate the differences in deformation structures between the two microstructures. A constitutive model was used to find that, surprisingly, twinning contributed relatively little to the strength of the alloy, chiefly due to the fine initial slip lengths that then gave rise to relatively little opportunity for work hardening by grain subdivision. Nevertheless, with lower high-cost alloying additions than equivalent Dual Phase steels (2-3 wt\% Mn) and greater ductility, medium-Mn TWIP$+$TRIP steels still represent an attractive area for future development.

\end{abstract}

\end{frontmatter}


\section{Introduction}

Medium Mn steels with Mn contents of 4-12 wt\% are seeing a rise in interest as a successor to high Mn (16-30 wt\% Mn) TWIP steels, named after the Twinning Induced Plasticity (TWIP) effect. TWIP steels are fully austenitic, have large elongations of $>$50\% and strain hardening rates of up to 3 GPa, resulting in a large amount of energy absorbed upon deformation \cite{DeCooman2018,Rahman2014,DeCooman2011,Allain2004a,Bouaziz2012}. However, due to the large Mn content, TWIP steels are unable to be processed using conventional secondary steelmaking processes \cite{Bausch2013,Bleck2007}. Additionally, the large Mn content has resulted in TWIP steels being significantly more expensive than other automotive strip steels \cite{Elliott2018}, \textit{e.g.} Dual Phase (DP) steels which are currently used in energy absorbing applications \cite{Billur2014a,Olsson2006}. A combination of these factors have led to limited industrial adoption of TWIP steels.

Within the past decade, many medium Mn steels have been developed with tensile properties that exceed those of TWIP steels. By reducing the Mn content, the microstructure of medium Mn steels are duplex ($\gamma + \alpha$). Depending on the thermomechanical processing history, a myriad of morphologies, phase fractions and distributions of the two phases can be produced. Regarding microstructure and morphology, the two most common microstructures are equiaxed and lamellar \cite{Han2017}. Medium Mn steels with equiaxed microstructures typically have polygonal and evenly distributed austenite and ferrite grains. Lamellar or laminated microstructures have alternating austenite and ferrite lamellae contained within a prior austenite grain. Most medium Mn steels are either equiaxed, lamellar or both \cite{Han2017,Ma2017,Li2019}. However, medium Mn steels with lamellar microstructures tend to be stronger, due to Hall-Petch strengthening arising from a smaller lamella thickness \textcolor{black}{although some exceptions exist} \cite{Han2014}, and are less prone to yield point elongation \cite{Sun2019a,Steineder2017}. Nevertheless, due to industrial thermomechanical processing limitations, it is difficult to produce a microstructure that is entirely equiaxed or lamellar. The  \textcolor{black}{second} medium Mn steel \textcolor{black}{to be} produced on an industrial scale by \textcolor{black}{v}oestalpine in Linz had a mixed lamellar and equiaxed microstructure \cite{Krizan2018,Krizan2019}.



In addition to the many types of microstructures that can be produced in medium Mn steels, many different plasticity enhancing mechanisms, on top of dislocation glide, can be activated in the austenite phase. By reducing the Mn content compared to high Mn TWIP steels, the stability of the austenite phase is lowered greatly such that stress-assisted or strain induced martensitic transformation is possible, \textit{i.e.} the Transformation Induced Plasticity (TRIP) effect. Additionally, if the Stacking Fault Energy (SFE) can be raised into the twinning regime (15-35 mJ m\textsuperscript{-2}) through element partitioning during an Intercritical Annealing (IA) heat treatment, the TWIP effect may also be activated \cite{Lee2014,Kwok2022b}. \textcolor{black}{Medium Mn steels can therefore also be grouped according to their plasticity enhancing mechanism, \textit{i.e.} TRIP-type or TWIP$+$TRIP-type \cite{Lee2015,Ma2017}.}

Many medium Mn steels have exhibited both TWIP and TRIP behaviour but they often occur in different austenite grains, \textit{e.g.} coarse austenite deforming \textit{via} TRIP and fine austenite deforming \textit{via} TWIP  \cite{Kwok2019,He2016,Lee2015d}. This means that the austenite grains do not deform homogeneously and such steels often exhibit serrations in the strain hardening curve. Therefore, the combined TWIP$+$TRIP effect, which occurs in a single austenite grain, is of particular interest as it allows for homogeneous deformation, sustained hardening during deformation and elongations in excess of 40\% \cite{Lee2014,Sun2018,Kwok2022b,Lee2015d,Sohn2014a,Sohn2017a}. However, the interplay between TWIP and TRIP can be very different. Lee \textit{et al.} \cite{Lee2014,Lee2015b} identified the successive TWIP$+$TRIP effect in a steel with austenite composition Fe-10.3Mn-2.9Al-2.0Si-0.32C. It was found that the austenite phase first formed twins and subsequently strain induced martensite at the twin intersections. This successive TWIP$+$TRIP effect led to a two-stage hardening behaviour where the first stage was twinning dominated and the second stage was transformation dominated. Sohn \textit{et. al.} \cite{Sohn2014a,Sohn2017a} seperately indentified a slightly different mechanism where twins and martensite formed concurrently and independently within an austenite grain during deformation. In their steel with austenite composition of Fe-11.5Mn-4.74Al-0.55C, the simultaneous TWIP$+$TRIP effect led to a multi-stage hardening behaviour and an exceptional 77\% total elongation.

Between the successive and simultaneous TWIP$+$TRIP mechanisms, the successive mechanism is more widely reported \cite{Sun2018, Lee2015e, Lee2014, Lee2015b} and is also the more commonly accepted mechanism of Strain Induced Martensite (SIM) nucleation and growth in Metastable Austenitic Stainless Steels (MASS) such as 304 and 301 \cite{Olson1972,Talonen2005,Tian2017}. The steel benefits from a high strain hardening rate ($>1.5$ GPa) because of two powerful plasticity enhancing mechanisms operating after each other. In the simultaneous mechanism, however, the twinning and transformation kinetics appear to be very slow, providing just enough hardening to avoid necking \cite{Sohn2014a}. Nevertheless, it allowed for a very steady engineering stress up to the failure strain of 77\%, signifcantly larger than medium Mn steels which exhibit the successive TWP$+$TRIP mechanism (40-50\%).

It is still uncertain what factors determine whether the austenite phase deforms \textit{via} successive or simultaneous TWIP$+$TRIP although factors such as microstructure, SFE and austenite stability are likely to play a large role \cite{Xu2017,Zhang2017a,Kwok2022b}. Furthermore, \textcolor{black}{the majority of} medium Mn steels shown to exhibit the TWIP$+$TRIP effect  \textcolor{black}{possess} equiaxed microstructures, \textit{i.e.} polygonal austenite grains. Not much is known how the TWIP$+$TRIP effect occurs in lamellar or mixed microstructures. This study therefore aims to explore the effect of microstructure on the interplay between TWIP and TRIP mechanisms and therefore their relative contributions. This would be achieved by producing two types of microstructures with similar austenite volume fraction and composition and examining how the various deformation structures evolve with strain.

\section{Experimental}

A steel ingot of dimensions 70 $\times$ 23 $\times$ 23 mm was produced \textit{via} vacuum arc melting. The bulk composition shown in Table \ref{tab:ICPcomp} was measured using Inductively Coupled Plasma (ICP) and Inert Gas Fusion (IGF). In order to simulate an industrial reheating cycle, the ingot was homogenised in a vacuum tube furnace at 1250 \degree C for 2 h and allowed to furnace cool to room temperature. \textcolor{black}{A previous study \cite{Kwok2022a} showed that the abovementioned homogenisation schedule was able to significantly reduce microsegregation in an arc melted ingot.} The ingot was then reheated to 1100 \degree C and rough rolled at the same temperature from 23 to 12 mm thickness in 4 passes and water quenched. The rough rolled ingot was then split along the long axis into two bars \textit{via} Electric Discharge Machining (EDM). The bars were then reheated to 1000 \degree C and finish rolled from 12 mm to 1.5 mm thick strip in \textcolor{black}{5} passes with a finish temperature of 850 \degree C. \textcolor{black}{The 5 passes were conducted at 50\%, 40\%, 30\%, 30\% and 25\% thickness reductions with a reheat between each pass.} One strip was water quenched immediately after the last pass, while the another strip was cooled to 600 \degree C and held for 30 min to simulate the coiling process and finally furnace cooled to room temperature. Both strips were then intercritically annealed at 750 \degree C for 10 min. The final samples produced either by furnace cooling or water quenching will be referred to as FC and WQ respectively. \textcolor{black}{In order to improve strain uniformity along the length of the rolled strip during finish rolling, the rolling speed was gradually increased after each pass to ensure a relatively constant rolling duration and minimise excessive temperature loss from the strip. However, it is acknowledged that under laboratory conditions, it is very difficult to guarantee precise temperatures and the degree of deformation across the length of each strip as well as between strips since the finish rolling of FC and WQ conditions were conducted separately.}

Tensile samples with gauge dimensions 19 $\times$ 3 $\times$ 1.5 mm were machined from both strips \textit{via} EDM, such that the tensile direction was parallel to the rolling direction. Tensile testing was subsequently conducted on an Instron load frame with a 30 kN load cell at a nominal strain rate of $10^{-3}$ s\textsuperscript{-1}.

Samples for Electron Backscattered Diffraction (EBSD) were mechanically ground and polished with an OPU polishing suspension. Samples for Transmission Electron Microscopy (TEM) and Transmission Kikuchi Diffraction (TKD) were prepared by mechanically thinning a 3 mm disk of material to below 90 \textmu m in thickness and subsequently electrolytic polished using a twin-jet electropolisher with a solution containing 5\% perchloric acid, 35\% butyl-alcohol and 60\% methanol at a temperature of $-40$ \degree C.

EBSD was conducted on a Zeiss Sigma FEG-SEM equipped with a Bruker EBSD detector. TEM and Energy Dispersive Spectroscopy (TEM-EDS) was conducted on on a JEOL JEM-F200 operated at an accelerating voltage of 200 kV. Transmission Kikuchi Diffraction (TKD) was conducted on a JEOL 7900 FEG-SEM.

A constitutive model based on the work by Lee \textit{et al.} \cite{Lee2014,Lee2015b,Lee2013b} and Latypov \textit{et al.} \cite{Latypov2016} was modified and used to model the plastic behaviour of the FC and WQ conditions. A full description of the model can be found in the Appendix and is available on GitHub \cite{Rose2022}. 

\section{Results}

\begin{table}[t]
	\centering
	\caption{Composition of the bulk steel in mass \% as measured by ICP and IGF for elements marked by $\dagger$.} 
	\begin{adjustbox}{width=\columnwidth,center}
		\begin{tabular}{cccccccc}
			\toprule
			Fe    & Mn    & Al    & Si    & C$^\dagger$     & N$^\dagger$     & P     & S$^\dagger$ \\
			\midrule
			Bal   & 4.77  & 2.75  & 1.51  & 0.505 & 0.003 & $<$0.005 & 0.002 \\
			\bottomrule
		\end{tabular}%
	\end{adjustbox}
	\label{tab:ICPcomp}%
\end{table}%

\subsection{Tensile properties}

\begin{figure}[t]
	\centering
	\includegraphics[width=\linewidth]{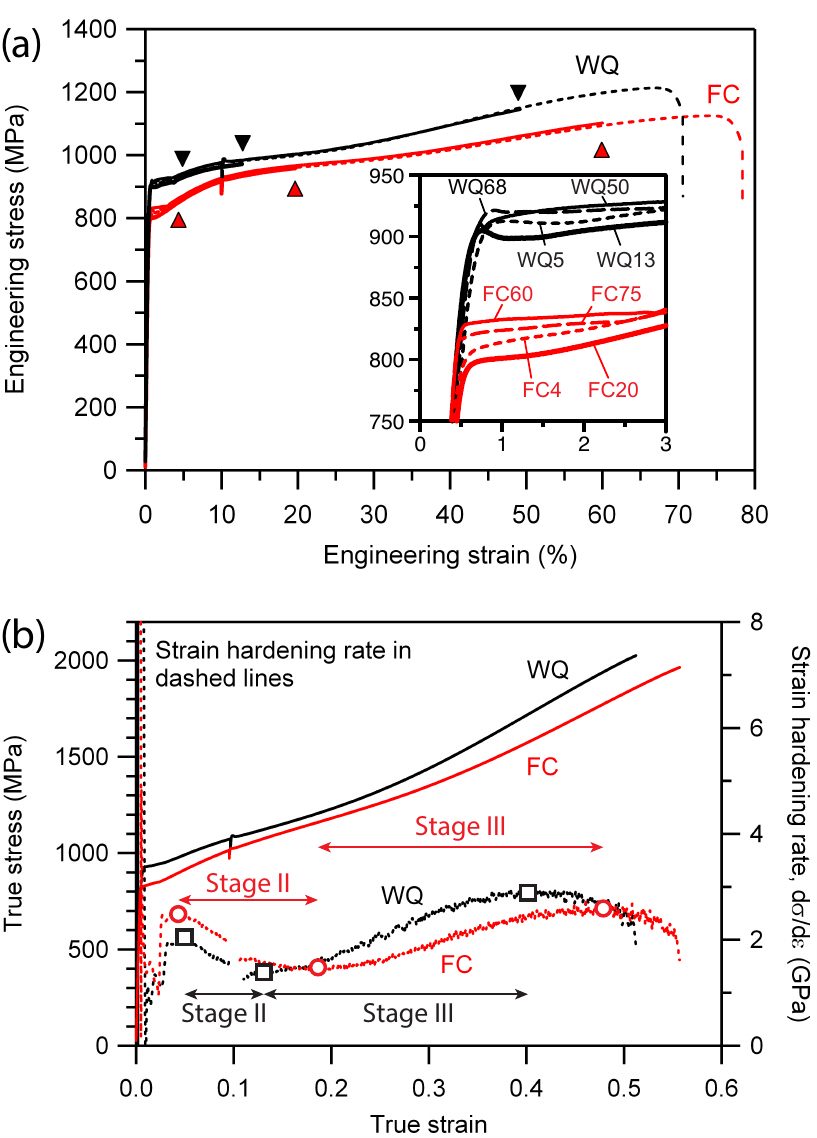}
	\caption{(a) Engineering stress strain curves \textcolor{black}{of all tensile specimens. Triangles indicate the strain to which the tensile test was interrupted}. Inset: early yielding behaviour \textcolor{black}{of all tensile samples}. (b) True stress strain curves and strain hardening rate drawn up to the point of uniform elongation \textcolor{black}{for tensile samples tested to failure}. Open red circles and black squares indicate true strains where interrupted tests were conducted. Stage I of the hardening rate was not labelled for clarity but refers to the true strain range between the onset of plastic deformation and the beginning of Stage II. N.B. extensometer was removed at 10\% \textcolor{black}{engineering} strain.}
	\label{fig:tensiles}
\end{figure}

\begin{figure*}[!t]
	\centering
	\includegraphics[width=\linewidth]{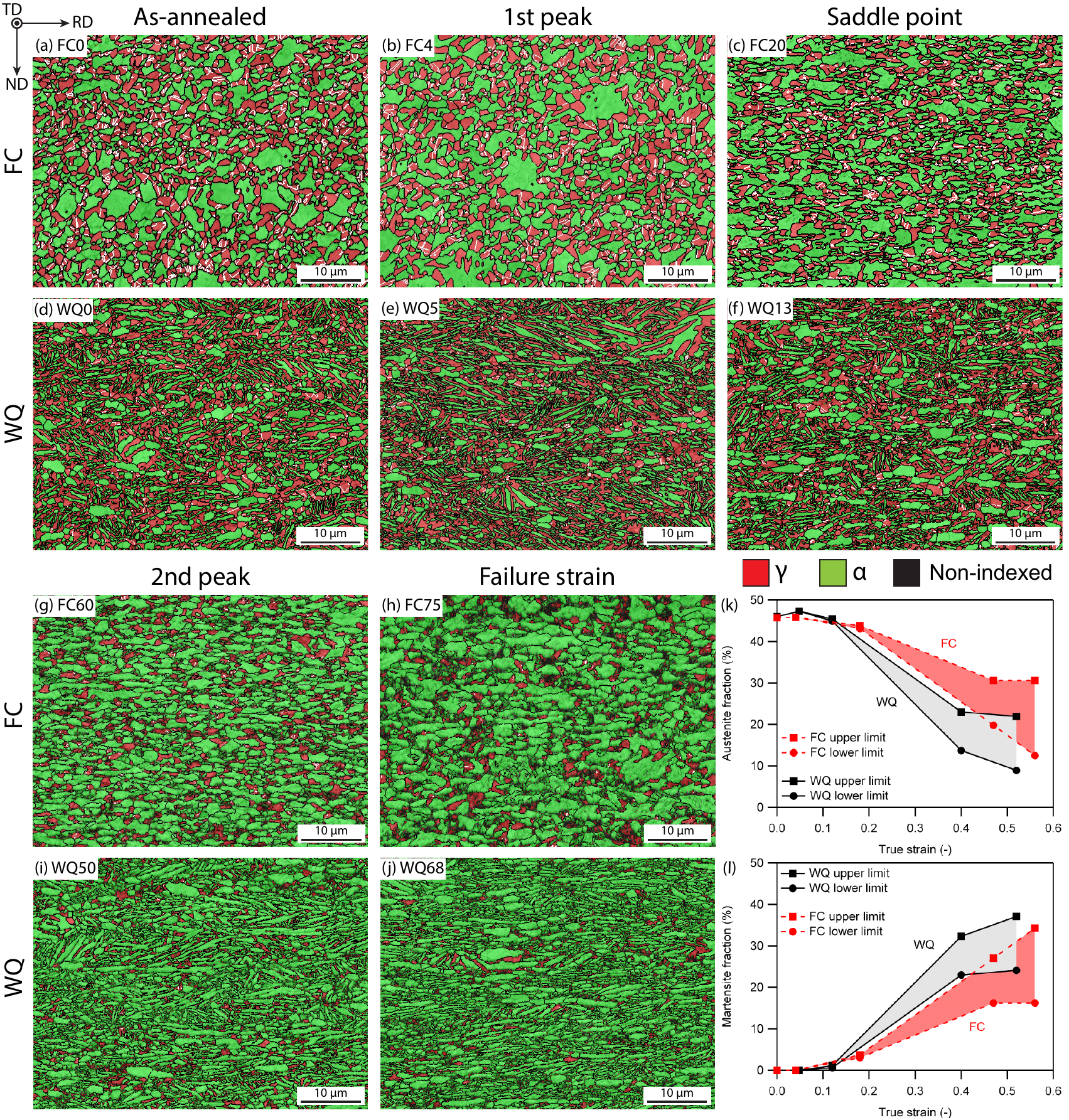}
	\caption{EBSD image quality and phase maps of (a) FC0, (b) FC4, (c) FC20, (d) WQ0, (e) WQ5, (f) WQ13, (g) FC60, (h)FC78, (i) WQ50, (j) WQ71. Red $-$ austenite, green $-$ ferrite/$\alpha'$-martensite, black $-$ non-indexed fraction. Black lines indicate High Angle Grain Boundaries (HAGBs) and white lines indicate austenite $\Sigma3$ boundaries, \textit{i.e.} annealing twin boundaries. (k) Summary of the change in austenite and (l) $\alpha'$-martensite phase fractions with true strain. \textcolor{black}{N.B. FC75 and WQ68 were obtained from the unifrom gauge section in the post mortem samples.}}
	\label{fig:ebsd-micros}
\end{figure*}

The tensile behaviour of the FC and WQ conditions are shown in Figure \ref{fig:tensiles} and the tensile properties are summarised in Table \ref{tab:tensiletable}. \textcolor{black}{Due to the small widths and thicknesses of the tensile samples, associated with the use of 400 g scale laboratory melts, the uncertainty budgets are greater than in full scale standards-certified tests \cite{Gabauer2000}, approximately $3\%$ for the yield stress. As a result, the yield and tensile strengths in Table \ref{tab:tensiletable} are given to two significant figures. The average yield strength could be determined from the various interrupted tensile samples; for the FC and WQ conditions, the average yield strengths were $810 \pm 15$ MPa (range 795 -- 829 MPa, n=4) and $910 \pm10$ MPa (range 894 -- 918 MPa, n=4) respectively, where n is the number of test samples. While there may be some variation in yield strengths due to the relative errors in the measurement of the tensile gauge cross section, Figure \ref{fig:tensiles}a showed that all tensile samples had very similar deformation behaviour, giving confidence in the uniformity of composition, strain and temperaure during finish rolling across the length of the rolled strip.} 

From Figure \ref{fig:tensiles}a, it can be seen that both FC and WQ steels have very high strengths ($>$800 MPa) and exceptional ductility ($>$70\%). The WQ sample had a higher yield and tensile strength but the FC sample had a slightly longer elongation. The inset in Figure \ref{fig:tensiles}a showed that \textcolor{black}{both FC and WQ samples} had a short yield point elongation of approximately 2\% strain.

When the Strain Hardening Rate (SHR) was obtained by differentiating the true stress-strain curve, it can be seen that they possessed a very similar shape, similar to many medium Mn steels \cite{Lee2014,Lee2015b,Lee2015c} and MASS \cite{Talonen2005,Talonen2007,Kovalev2011,Choi2014} that exhibit the successive TWIP$+$TRIP effect. \textcolor{black}{Three} hardening stages can be identified. At stage I, \textit{i.e.} the onset of plastic deformation, both SHR curves showed a rapid decrease, reaching a local minima at approximately 0.02 true strain before \textcolor{black}{increasing} rapidly to the first peak between 0.04 and 0.05 true strain. At stage II, the SHR then decreases \textcolor{black}{from the first peak} to a saddle point and at stage III it rises again slowly to the second peak. While the shape of the SHR curves may be similar, the strain at which the first peak, saddle point and second peaks occur as well as its value differ slightly between the FC and the WQ steel. To investigate the microstructural evolution at these three unique points \textcolor{black}{(first peak, saddle point and second peak)}, interrupted tensile tests were conducted up to the corresponding strains for the FC and WQ steel. The as-annealed and interrupted tensile samples are henceforth named FC0, FC4, FC20, FC60 for the FC steel and WQ0, WQ5, WQ13, WQ50 for the WQ steel where the digits represent the engineering strain \textcolor{black}{(not true strain)} to which they were tested, rounded to the nearest percent.

\begin{table}[t!]
	\centering
	\caption{Summary of tensile properties \textcolor{black}{of the tensile samples tested to failure.} E - Young's modulus, $\sigma_{0.2}$ - 0.2\% proof stress, MPa\% \textcolor{black}{determined as the product of UTS and $\varepsilon$}.}
	\begin{tabular}{lccccc}
		\toprule
		&	E	&	$\sigma_{0.2}$	&	UTS	& $\varepsilon$  & MPa\% \\
		& (GPa)	&	(MPa)	&	(MPa)	&	(\%)	&	($\times10^3$)\\
		\midrule
		FC    & 210  & 820   & 1100  & 78	& 86\\
		WQ    & 190  & 900   & 1200  & 71	& 85\\
		\bottomrule
	\end{tabular}%
	\label{tab:tensiletable}%
\end{table}%

\begin{figure}[t]
	\centering
	\includegraphics[width=\linewidth]{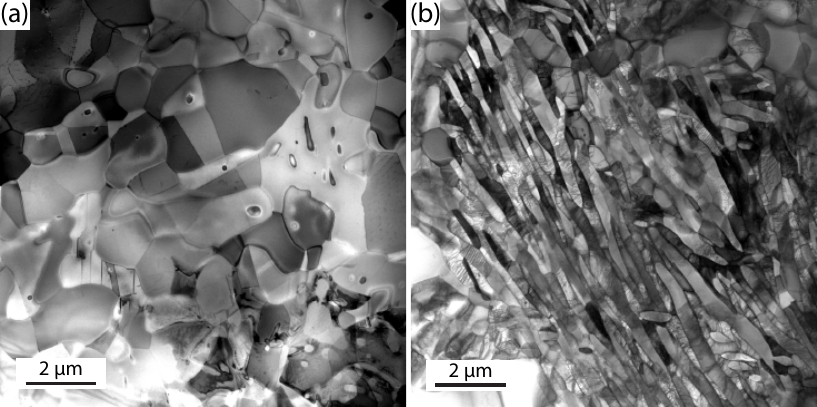}
	\caption{STEM-BF images of (a) FC and (b) WQ in the as-annealed condition. }
	\label{fig:as-annealed}
\end{figure}

\begin{figure*}[p!]
	\centering
	\includegraphics[width=\linewidth]{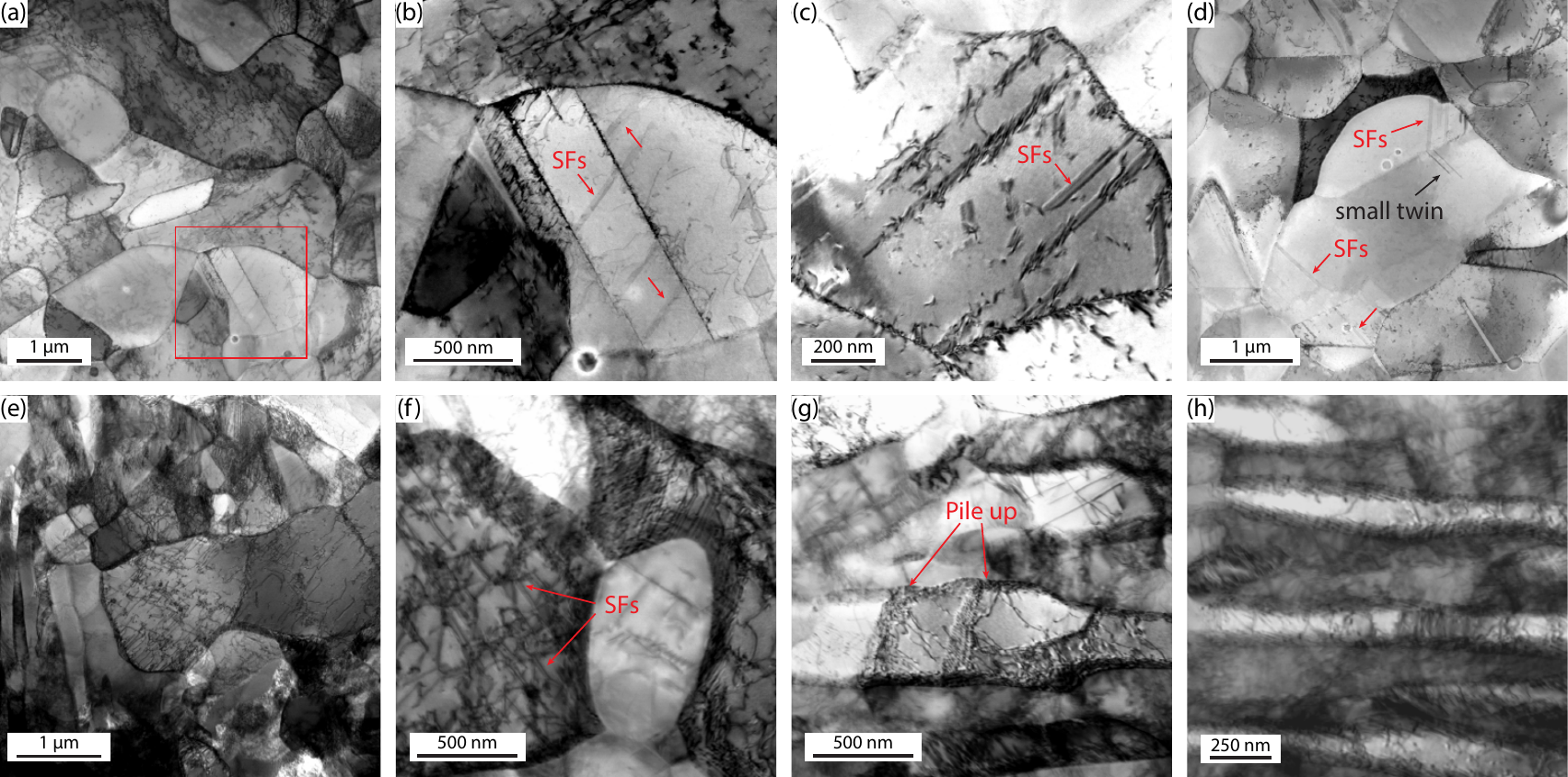}
	\caption{STEM-BF of (a-d) FC4 and (e-h) WQ5 at the first SHR peak. (a) General microstructure of FC steel, (b) magnified view of red square in (a) showing stacking faults inside an austenite grain and its annealing twin. (c) Stacking faults in a separate austenite grain. (d) Stacking faults (red arrows) emitted from a grain boundary terminating at a subgrain boundary and small twins (black arrow) emitted from the subgrain boundary. (e) General microstructure of WQ steel, (f) partial dislocations within an equiaxed austenite grain. (g) Dislocation pile up across a lamellar grain. (h) Dislocations emitted from interphase boundaries of lamellar grains.}
	\label{fig:firstbumptem}
\end{figure*}

\subsection{Microstructure}

EBSD Image Quality and Phase Maps (IQ+PM) of the FC and WQ steels at varying strain are shown in Figure \ref{fig:ebsd-micros}. In the as-annealed condition, FC0 showed a largely equiaxed microstructure (Figure \ref{fig:ebsd-micros}a). It was likely that the cementite in the pearlite matrix globularised and transformed into austenite during the IA, while WQ0 adopted a mixed equiaxed $+$ lamellar microstructure. The lamellar regions arose when austenite nucleated in between martensite laths, while the equiaxed regions formed due to some degree of recrystallisation of martensite, predominantly around the PAGBs.

With increasing strain, it was observed that the austenite fraction began to decrease after the first peak, indicating that the TRIP effect was operative in both FC and WQ steels. However, due to the increasing Non-Indexed (NI) fraction at higher strain levels, there was some uncertainty in the calculated martensite fraction, which indexes as BCC in EBSD. It is acknowledged that the NI regions are typically either martensite or deformed austenite. Therefore, an upper and lower limit was established in Figures \ref{fig:ebsd-micros}k-l where the upper martensite limit was obtained if the entire NI fraction was treated as martensite and the lower martensite limit was obtained if the entire NI fraction was treated as austenite and \textit{vice versa} for the austenite upper and lower limits. \textcolor{black}{In Figure \ref{fig:ebsd-micros}k, it appears that the austenite fraction increased slightly between WQ0 and WQ5. This was likely due to a small variation in microstructure between the tensile specimens as austenite fraction cannot \textcolor{black}{increase} with strain.}

Between the as-annealed condition and the first peak, there was no transformation in both FC and WQ steels, suggesting that an incubation strain was needed for the TRIP effect. Subsequently, with increasing strain, the martensite fraction in the WQ steel increased at a higher rate compared to the FC steel with a larger final fraction at the failure strain. In the FC steel, both ferrite and austenite grains were observed to elongate in the tensile direction while in the WQ steel, the lamellar regions were also observed to elongate but additially rotating and orienting themselves parallel to the tensile direction at the failure strain. 


While EBSD was able to provide macroscopic insights into the microstructural evolution, it was necessary to probe the finer deformation structures using TEM. Figure \ref{fig:as-annealed} shows the microstructure of the FC and WQ steel in the as-annealed condition under STEM-BF. The as-annealed FC microstructure showed an equiaxed microstructure with a low dislocation density. On the other hand, the as-annealed WQ microstructure showed a lamellar microstructure with average lamella width of 290 nm. A relatively larger number of dislocations were observed in the WQ microstructure. Han \textit{et al.} \cite{Han2014} also found that the ferrite phase in lamellar type microstructures had a higher dislocation density than equiaxed ferrite due to the lack of recrystallisation during IA. While the FC microstructure was not obtained \textit{via} cold rolling and recrystallising, the multiple phase transformations during coiling, furnace cooling and IA were likely able to eliminate any residual dislocation density after hot rolling.

\subsubsection{Stage I: zero strain to first peak}
\textcolor{black}{At stage I, there was no transformation as shown in Figure \ref{fig:ebsd-micros}k.} From Figure \ref{fig:firstbumptem}, it was observed that dislocation multiplication was occuring in the austenite and ferrite phases in both the FC and WQ conditions. \textcolor{black}{Twinning in both conditions was also very limited.} In FC4, multiple Stacking Faults (SFs) were observed growing from austenite and annealing twin boundaries. In WQ5, SFs were also observed but only in the more globular austenite grains. In the lamellar grains however, no SFs were observed. Instead, dislocations were emitted from the interphase boundaries and also seen to be piling up across the width of certain lamellar grains. The higher density of SFs in FC4 may also explain why the SHR at the first peak was higher than WQ5. 


In many fine grained equiaxed medium Mn steels, the first peak is the result of yield point elongation \cite{Lee2015b,Lee2015e}. Sun \textit{et al.} \cite{Sun2019a} have shown that this was because of a rapid dislocation generation from the large amount of $\gamma/\alpha$ interfaces in a relatively dislocation free microstructure. Lamellar microstructures typically exhibit continuous yielding due to a higher dislocation density within the lamellar grains, as seen in Figure \ref{fig:as-annealed}b. However, because WQ was a mixed microstructure, combining both equiaxed and lamellar regions, a short yield point elongation was still present.



\subsubsection{Stage II: first peak to saddle point}

\begin{figure*}[p!]
	\centering
	\includegraphics[width=\linewidth]{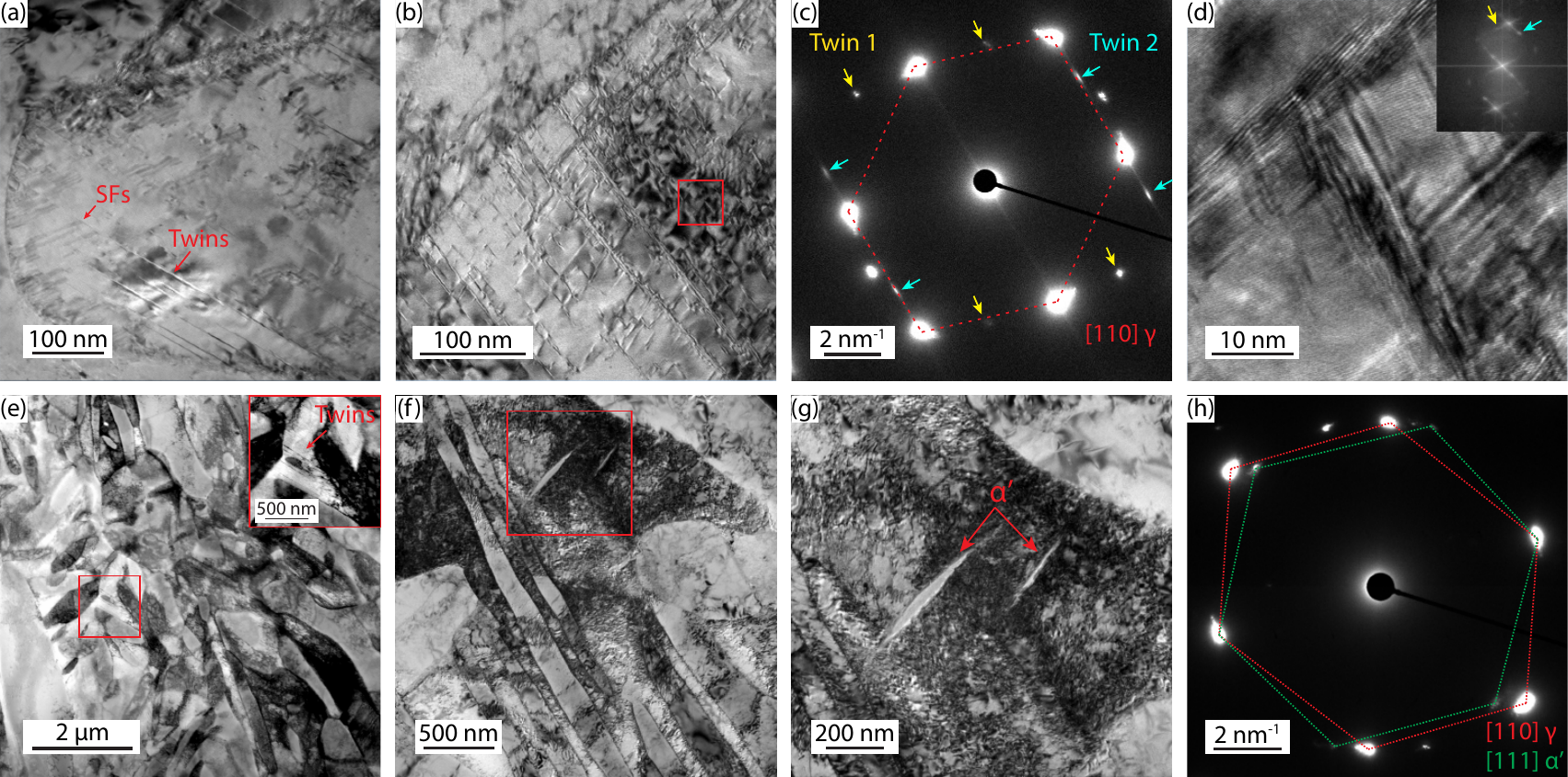}
	\caption{TEM micrographs of (a-d) FC20 and (e-h) WQ13 steels at their respective SHR saddle points. TEM-BF of (a) twins in one twinning direction and SFs in the other direction, (b) two twinning systems in an austenite grain. (c) Diffraction pattern obtained from the austenite grain in (b) showing twinning spots from two systems. (d) HR-TEM of red square in (b) showing twin intersection. Inset: pseudo-diffraction pattern of $[110]_\gamma$ showing twinning spots. (e) STEM-BF mmicrograph of the general microstructure of WQ steel with very limited twinning. (f) TEM-BF micrograph of thin $\alpha'$-martensite laths nucleating in deformed austenite. (g) TEM-BF micrograph of the red square in (f). (h) Diffraction pattern obtained from (g) showing austenite and $\alpha'$-martensite with the KS-OR. }
	\label{fig:firstsaddletem}
\end{figure*}

Figure \ref{fig:firstsaddletem} shows the microstructures of FC20 and WQ13 at the SHR saddle point. In FC20, some austenite grains were observed to have one set of twins, while others were observed to have two. Figure \ref{fig:firstsaddletem}a shows an austenite grain with twins growing out from the grain boundary while stacking faults were growing in the other twinning direction. The SFs appear to have either nucleated from the grain boundary or from the first twins. In another austenite grain in FC20 as shown in Figures \ref{fig:firstsaddletem}b-c, two twinning systems were clearly operating as also shown in the diffraction pattern where two sets of twinning spots were observed. According to the successive TWIP$+$TRIP mechanism described by Lee \textit{et al.} \cite{Lee2014,Lee2014a}, $\alpha'$-martensite at twin intersections should be observed at this strain. However, additional diffration spots associated with martensite transformation were not observed at the twin intersections in FC20. High Resolution TEM (HR-TEM) of the twin intersections also showed no martensite at the twin intersections. 


In WQ13, there was very limited twinning compared to FC20. In Figure \ref{fig:firstsaddletem}e, only a small twinned region could be found and was not located in a lamellar region. In Figures \ref{fig:firstsaddletem}f-g, several thin laths of martensite were found in an austenite grain adjacent to a lamellar region. The diffraction pattern in Figure \ref{fig:firstsaddletem}h showed that the martensite laths had a Kurdjumov-Sachs orientation relationship (KS-OR) with the parent austenite. Thin martensite laths were also observed by Lee \textit{et al.} \cite{Lee2015c} in a lamellar medium Mn steel.

\subsubsection{Stage III: saddle point to second peak}

\begin{figure*}[t]
	\centering
	\includegraphics[width=\linewidth]{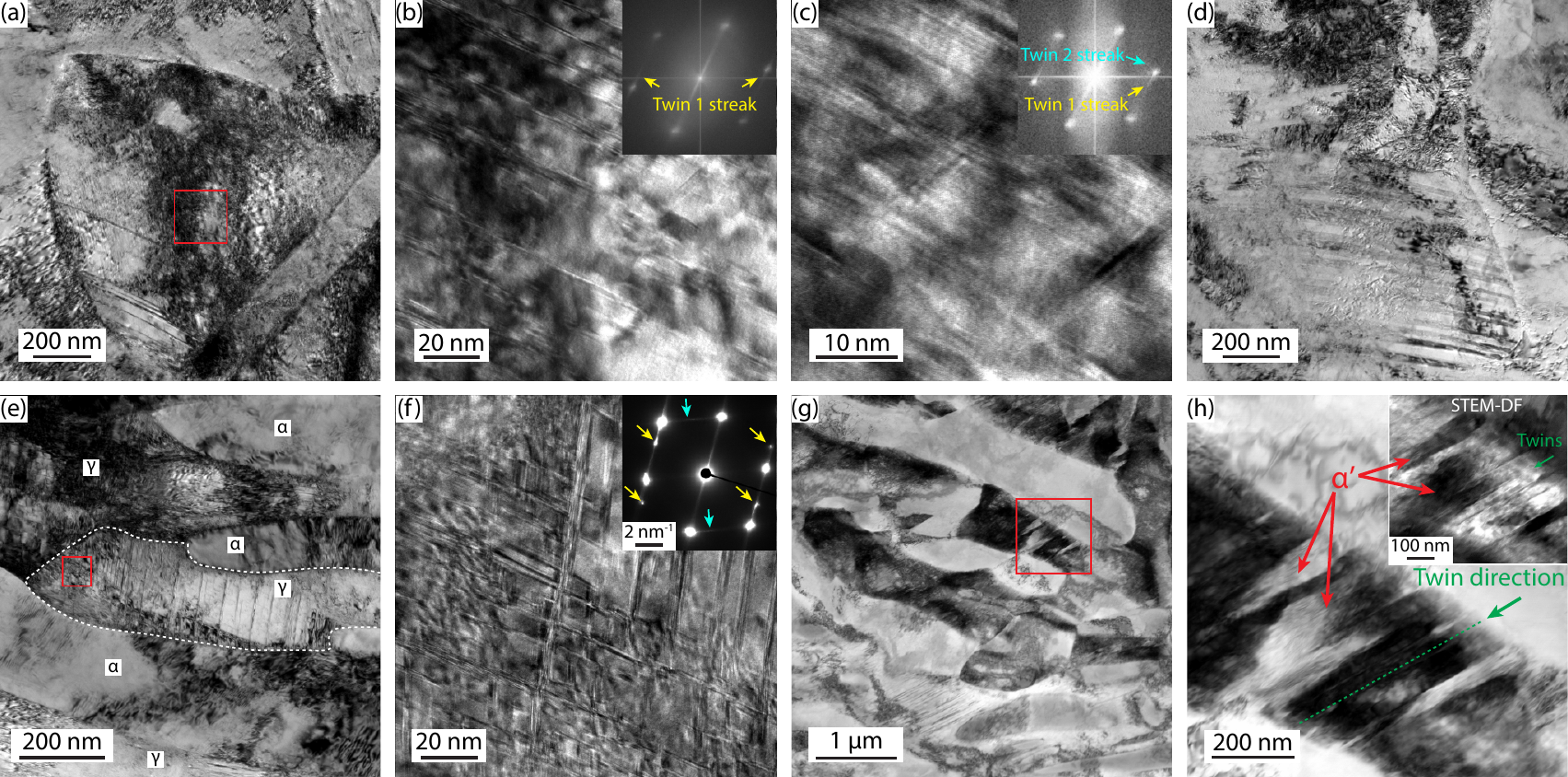}
	\caption{TEM micrographs of (a-d) FC60 and (e-h) WQ50 at their respective second SHR peaks. TEM-BF of (a) a deformed austenite grain demonstrating long twins near the bottom. HR-TEM micrographs of (b) magnified area of the red square in (a) demonstrating shorter and thinner twins, inset: pseudo-diffraction pattern of $[110]_\gamma$ demonstrating twinning streaks, (c) further magnification of (b) revealing short twins from the second twinning system, inset: pseudo-diffraction pattern of $[110]_\gamma$ demonstrating two twinning streaks. (d) TEM-BF of an austenite grain showing thick twins. TEM-BF of (e) lamellar region in WQ50 showing heavily twinned austenite grains and the highlighted austenite grain having a larger twin density at the lamella tip, (f) magnified image from red square in (e) showing two active twinning systems. Inset: diffraction pattern of the twinned region. Yellow arrows point to primary twin spots, cyan arrows point to secondary twin streaks. Beam parallel to $[110]_\gamma$. STEM-BF of (g) another lamellar region and (h) magnified image from red square in (g) showing $\alpha'$-martensite growing across a twinned austenite lamella in the twinning direction. Inset: STEM-DF of (h) highlighting the twins.    }
	\label{fig:secondpeaktem}
\end{figure*}

At the second SHR peak, the TEM micrographs of FC60 and WQ50 are shown in Figure \ref{fig:secondpeaktem}. In FC60, many twinned austenite grains could be observed such as the one shown in Figure \ref{fig:secondpeaktem}a. With increased magnification. Much finer but shorter twins were observed within the grain interior (Figure \ref{fig:secondpeaktem}b) and when the magnification was increased further, several short twins from the second twinning system could be observed (Figure \ref{fig:secondpeaktem}c). This suggests that twinning continued to occur even up to 60\% strain in the FC steel. In another austenite grain as shown in Figure \ref{fig:secondpeaktem}d, twin thicknening was observed which has often been found in TWIP steels at large strains \cite{Mahajan1973} and was also observed by Sohn \textit{et al.} \cite{Sohn2014a} in a medium Mn steel.

In WQ50, a large number of twinned lamellar austenite grains were observed. This suggests that in WQ, twinning was more active in stage III compared to stage II. Figure \ref{fig:secondpeaktem}a shows a lamellar austenite grain with twins that grew across the grain. Further magnification revealed secondary twins at the tip of the lamellar grain. However, the diffraction pattern from the tip still showed that there was no $\alpha'$-martensite at the twin intersections. Nevertheless, in another grain (Figures \ref{fig:secondpeaktem}g-h), martensite laths were observed growing from the grain boundary across a twinned lamellar austenite grain in the same direction as the twins.

\begin{figure*}[t]
	\centering
	\includegraphics[width=\linewidth]{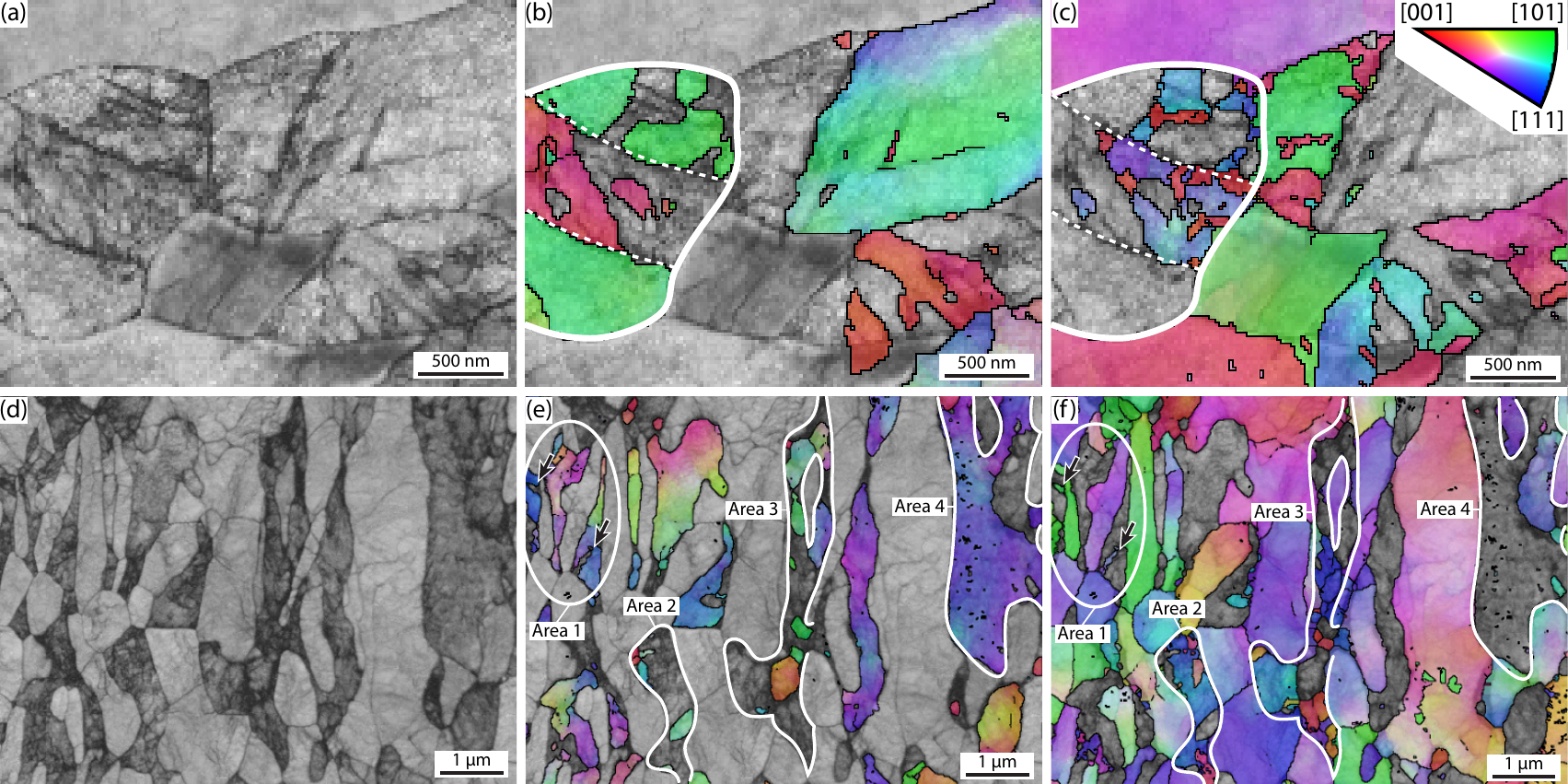}
	\caption{TKD micrographs of (a-c) FC60 and (d-f) WQ50. (a) Band contrast, (b) FCC IPF-Z map and (c) BCC IPF-Z map of FC60 sample. (d) Band contrast, (e) FCC IPF-Z map and (f) BCC IPF-Z map in WQ50 sample. Z-direction points out of the page. Black lines indicate HAGBs.}
	\label{fig:secondpeaktkd}
\end{figure*}

While TEM has effectively revealed the twinned structures in both FC and WQ steels, it was difficult to identify martensitic regions. For this reason, TKD was conducted on the TEM foils from FC60 and WQ50. The resulting data \textcolor{black}{are} shown in Figure \ref{fig:secondpeaktkd}. In the FC60 sample (Figures \ref{fig:secondpeaktkd}a-c), a deformed equiaxed austenite grain with a curved annealing twin (outlined in white) could be observed from the Band Contrast (BC) and FCC IPF-Z maps. From the BCC IPF-Z map (Figure \ref{fig:secondpeaktkd}c), BCC regions were observed within the outlined austenite grain. These BCC grains were observed to be within $5\degree$ of the KS-OR with the parent austenite grain and were therefore different variants of $\alpha'$-martensite. These blocky $\alpha'$-martensite grains appear to have first nucleated from the austenite grain boundaries and then from each other, suggesting that $\alpha'$-martensite growth was limited and nucleation may have occured continuously with increasing strain. The limited growth may explain why size of the blocky $\alpha'$-martensite grains remained very small. However, it cannot be said that the $\alpha'$-martensite observed in Figure \ref{fig:secondpeaktkd} only formed during stage III, it is likely that some $\alpha'$-martensite also formed during stage II. 

In the WQ50 sample (Figures \ref{fig:secondpeaktkd}d-f), the austenite grains before transformation \textcolor{black}{can be identified} as the dark grey \textcolor{black}{regions}, \textit{i.e.} more deformed regions in the BC map (Figure \ref{fig:secondpeaktkd}d). This is also confirmed in Figure \ref{fig:secondpeaktkd}e where the untransformed austenite lie within the dark grey regions. In Figures \ref{fig:secondpeaktkd}e-f, four areas of interest are highlighted. In area 1,  the arrows point to $\alpha'$-martensite grains with a lath morphology which were observed growing across austenite lamellae as observed in WQ13 (Figures \ref{fig:firstsaddletem}f-h). However, this morphology was an exception rather than the rule as similar lath martensite morphologies were not commonly observed elsewhere. It is also worth noting that the three austenite lamellae contained in area 1 remained largely untransformed even up to 50\% strain. In area 2, the outlined region was likely a globular prior austenite grain but was elongated at 50\% strain. The prior austenite grain was mostly transformed with only a few pockets of austenite left. In the BCC IPF-Z map in Figure \ref{fig:secondpeaktkd}f, several small martensite grains were observed at the prior austenite grain boundary in a similar manner as described in FC60. However, the prior austenite grain was mostly dominated by a single martensite grain with $[111]$ direction out of the page. In area 3, the outlined prior austenite grain was similarly partially transformed to martensite. In the more bulbous region near the bottom left, a single martensite grain with $[111]$ direction out of the page also dominated the region. However, at the bottom tip and along the length of the lamellar grain, the austenite grain transformed into relatively equally sized submicron martensite grains. Finally, in area 4, the untransformed austenite grain contained several extremely fine intragranular martensite grains. These martensite grains were significantly finer and do not appear to have nucleated from a grain boundary in the same manner as in the aforementioned three areas. It is therefore possible that these fine martensite grains are of the twin-twin intersection variety which forms in the successive TWIP$+$TRIP mechanism.



Finally, \textcolor{black}{beyond stage III}, comparing the EBSD phase maps in Figures \ref{fig:ebsd-micros}g-l, the austenite phase still continued to transform to martensite. However, \textcolor{black}{the strain region beyond stage III} is characterised \textcolor{black}{by} decreasing SHR (Figure \ref{fig:tensiles}b), suggesting that both twin and martensite fractions were both approaching saturation until both steels failed by necking, \textit{i.e.} $\sigma = d\sigma/d\varepsilon$.

\subsection{Composition}

The composition of austenite and ferrite in both FC and WQ steels were measured using TEM-EDS. The results are shown in Table \ref{tab:TEM-EDScomp}. The C content in austenite was determined by using the lever rule, assuming negligible C solubility in ferrite. \textcolor{black}{It should be noted that the bulk composition as measured by TEM-EDS, \textit{i.e.} $X_\gamma  V_f^\gamma + X_\alpha V_f^\alpha$ where $X_i$ is the Mn, Al or Si content and $V_f^i$ is the volume fraction of phase $i$, may not be the same as the bulk composition as measured using ICP. The difference may be attributed to the limitations of quantitative TEM-EDS. However, the resolution of TEM-EDS was needed to probe the compositions of the fine grained microstructures in the FC and WQ samples.} The SFE was calculated according to the method proposed by Sun \textit{et al.} \cite{Sun2018}. Md\textsubscript{30}, defined as the temperature where half of the austenite transforms to martensite at a strain of 30\% was calculated according to the following equation\cite{Angel1954,Nohara1977}:

\begin{equation}
\begin{split}
\small
Md_{30} (\degree C)= \,& 551 - 462 C  - 8.1 Mn  - 9.2 Si \\
& - 1.42(-3.29-6.64\log_{10}d_\gamma-8)
\end{split}
\end{equation}

\noindent
where compositions are given in mass \% and $d_\gamma$ is the austenite grain size in \textmu m. A higher Md\textsubscript{30} indicates lower austenite stability against strain induced martensitic transformation and \textit{vice versa}. The Ms temperature, defined as the temperature where athermal martensite begins to form upon rapid cooling from austenite, was calculated according to the following equation \cite{Lee2013d}:

\begin{equation}
\label{eq:Ms}
\small
Ms (\degree C)= 475.9-335.1 C - 34.5Mn - 1.3 Si + 11.67\ln (d_\gamma)
\end{equation}

\noindent
A higher Ms indicates lower austenite stability against athermal martensitic transformation and \textit{vice versa}. Both Md\textsubscript{30} and Ms have been used to qualitatively determine the stability of austenite against strain induced martensitic transformation.

\begin{table*}[t!]
	\centering
	\caption{TEM-EDS measurements in mass percent of alloying elements in each phase in the FC0 and WQ0 samples, balance being Fe. Standard error given in paranthesis. $d$\textsubscript{grain} calculated as the equivalent circle diameter from EBSD, except for * which is the average austenite lamellar width in WQ0. N.B. the morphology of the grains in FC0 and WQ0 samples are equiaxed and lamellar respectively. $\dagger$ C content determined by lever rule assuming negligible C content in ferrite. \textcolor{black}{$V_f$ - volume fraction.}}
	\small
	\begin{tabular}{lccccccccc}
		\toprule
		& Mn    & Al    & Si    & C$^\dagger$    & $V_f$ & $d$\textsubscript{grain} & SFE   &  Ms  &  Md\textsubscript{30} \\
		& \multicolumn{4}{c}{}    &       & (\textmu m) & (mJ m$^{-2}$) & (\degree C) & (\degree C) \\
		\midrule
		\smallskip
		Bulk (ICP) & 4.77  & 2.75  & 1.41  & 0.505 & -    & -    & -    & -    & - \\
		$\gamma$ FC0  & 8.40 (0.38) & 1.07 (0.17) & 1.00 (0.11) & 1.10 & 0.46  & 1.5   & 34  & -179 & -17\\
		
		$\gamma$ WQ0  & 7.30 (0.60) & 1.80 (0.10) & 1.00 (0.11) & 1.07 & 0.47  & 1.2 & 31 &  -133  &  5 \\
		\smallskip
		&					&						&					&		&			&	0.3*	&	31&	-150	&	-1\\
		$\alpha$ FC0  & 3.81 (0.10) & 2.46 (0.04) & 1.46 (0.05) & 0     & 0.54  & 3.0     &   -    &   -    & -  \\
		$\alpha$ WQ0  & 3.92 (0.05) & 2.25 (0.12) & 1.37 (0.07) & 0     & 0.53  & 1.3   &  -     &   -    & - \\
		\bottomrule
	\end{tabular}%
	\label{tab:TEM-EDScomp}%
\end{table*}%

Comparing the austenite compositions between the FC and WQ states in Table \ref{tab:TEM-EDScomp}, it can be seen that the FC condition had a slightly higher Mn content while the other elements remained relatively equal. This may be attributed to the additional coiling and furnace cooling steps during the processing of the FC condition which provided additional time for Mn to partition out of ferrite and into the cementite phase. During the IA step, the cementite then globularised and transformed into austenite with an enriched Mn content compared to the WQ condition. The difference in Mn content resulted in a slightly higher SFE and lower Md\textsubscript{30} and Ms for FC. 

The SFE of FC and WQ steels were both within the predicted twinning regime of TWIP steels \cite{DeCooman2018} and also medium Mn steels \cite{Kwok2022b}. On the other hand, because of the large C content in both steels ($>1$ wt\%), the Md\textsubscript{30} and Ms temperatures were very low, suggesting that the austenite was very stable. In medium Mn steels, Lee \textit{et al.} \cite{Lee2014} demonstrated that it was possible to overstabilise the austenite phase such that the TRIP effect no longer becomes operative. While the TRIP effect was clearly observed in both FC and WQ, the high austenite stability would certainly have an effect on the nucleation and growth of $\alpha'$-martensite grains.

\subsection{Modified constitutive model}



\begin{figure*}[t!]
	\centering
	\includegraphics[width=\linewidth]{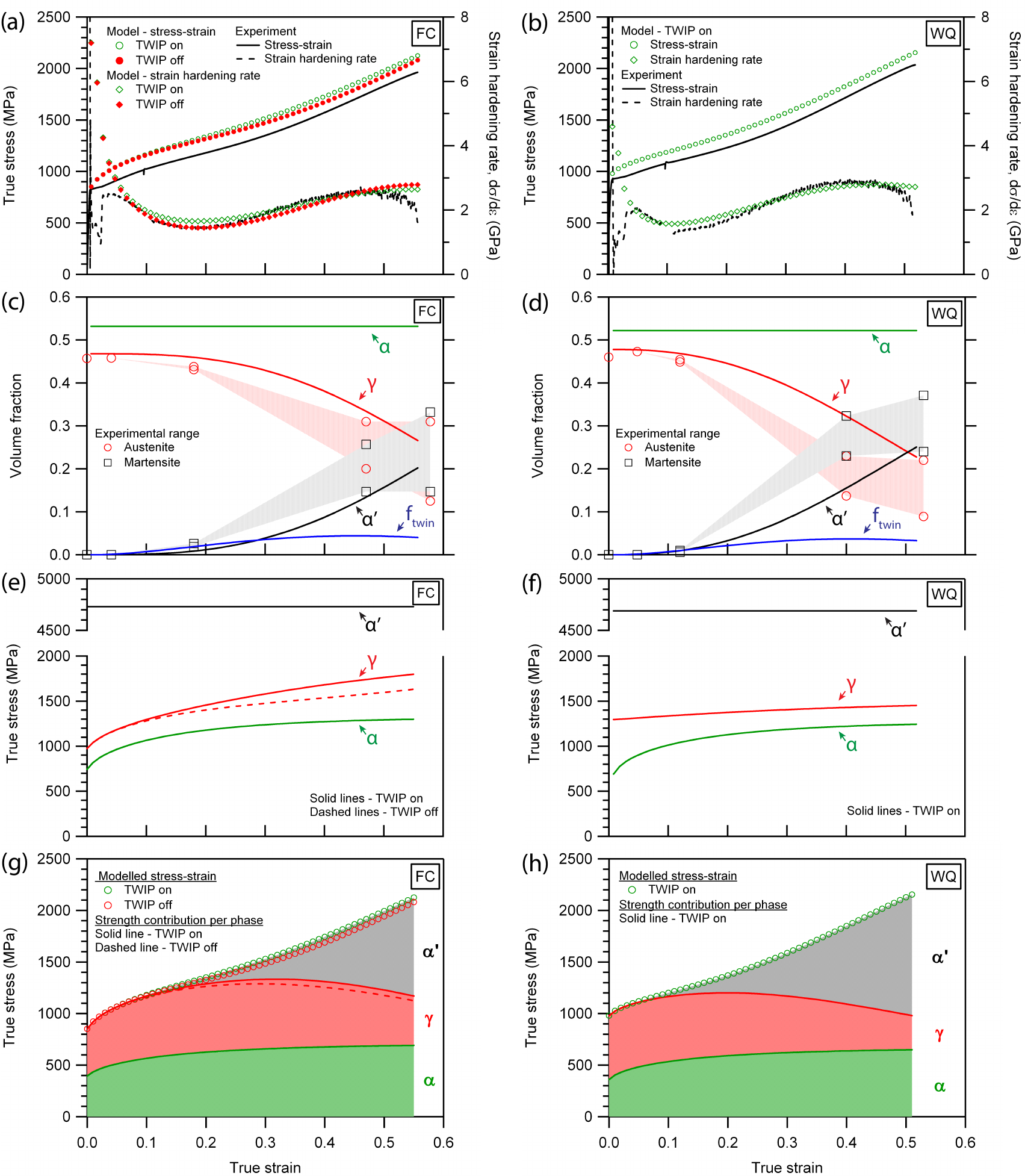}
	\caption{Experimental and modelled tensile curves of the (a) FC and (b) WQ conditions.  Experimental and modelled changes in phase fraction with strain of the (c) FC and (d) conditions. Twin fraction, f\textsubscript{twin}, refers to the fraction of twins in austenite multiplied by the austenite fraction. Component stress-strain curves of the individual phases of the (e) FC and (f) WQ conditions. Cumulative contributions to global strength from the individual phases in the (g) FC and (h) WQ conditions. N.B. Modelled TWIP off curves for the WQ condition were not shown for clarity as they were nearly identical to TWIP on curves. TWIP on and off curves are very similar in FC condition and nearly identical in WQ condition due to limited dislocation storage ability at slip lengths on the micron and submicron level.}
	\label{fig:fc-and-wq-modelling-combined}
\end{figure*}


After \textcolor{black}{examining} the microstructural evolution with strain using EBSD, TEM and TKD, it \textcolor{black}{is} evident that \textcolor{black}{it was} the simultaneous, \textcolor{black}{rather than the successive} TWIP$+$TRIP mechanism \textcolor{black}{that} was active in both FC and WQ conditions as $\alpha'$-martensite grains were not observed at twin intersections and have mostly nucleated at austenite grain boundaries. \textcolor{black}{Microstructural examination also showed that the evolution of deformation structures such as twins and $\alpha'$-martensite were very different between FC and WQ conditions. The different twinning kinetics theoretically should have resulted in a clear difference in strain hardening profiles between the FC and WQ conditions. However, from Figure \ref{fig:tensiles}b, the strain hardening rate curves between FC and WQ conditions showed a very similar profile.}

\textcolor{black}{In order to reconcile the seemingly conflicting observations between the tensile properties and microstructure evolution, a constitutive model developed by Lee \textit{et al.} \cite{Lee2014,Lee2015b,Lee2013b} was used to determine if the tensile properties in Figure \ref{fig:tensiles} could be reproduced given the microstructural data from Figures \ref{fig:ebsd-micros}-\ref{fig:secondpeaktkd} and \textit{vice versa}. Since the constitutive model was initially developed for medium Mn steels that exhibited the successive TWIP$+$TRIP mechanisms, two key changes were made in order to accomodate the simultaneous TWIP$+$TRIP mechanism. Firstly, equations responsible for the evolution of martensite fraction with strain were replaced with a single Avrami equation \cite{Avrami1939,Avrami1940,Avrami1941} which also effectively uncouples the dependence of TRIP on TWIP. Secondly, following the findings of Latypov \textit{et al.} \cite{Latypov2016}, the strength of the $\alpha'$-martensite phase was approximated to be constant with strain. The results from the modified constitutive model are shown in Figure \ref{fig:fc-and-wq-modelling-combined}.}

From Figure \ref{fig:fc-and-wq-modelling-combined}a, a good agreement between the model and experimental stress-strain and SHR curves was observed for the FC condition. A slight overprediction was observed in the modelled stress-strain curve but can be attributed to the model assuming continuous yielding and not accounting for the slight yield point elongation in the FC condition. From Figure \ref{fig:fc-and-wq-modelling-combined}c, the predicted austenite and $\alpha'$-martensite phase fractions were shown to be in reasonable agreement with the experimental ranges as determined using EBSD in Figure \ref{fig:ebsd-micros}. This largely confirms the validity of the modified constitutive model for modelling the simultaneous TWP$+$TRIP mechanism in equiaxed-type microstructures. 

However, the same model did not work as well when applied to the WQ condition. In Figure \ref{fig:fc-and-wq-modelling-combined}b the model appears to show a good fit with the SHR curve, however the predicted $\alpha'$-martensite fraction fell short of the experimentally determined range at a true strain of 0.4 in Figure \ref{fig:fc-and-wq-modelling-combined}d. This was largely attributed to the mixed equiaxed and lamellar grain morphology in the WQ condition \textcolor{black}{which might have led to a more complex strain partitioning mechanism that might not be best represented with the current equations.} Nevertheless, the model was able to reasonably predict the initial and final martensite fractions in the WQ condition.

Since TWIP and TRIP mechanisms have been uncoupled in the modified constitutive model, it is possible to model the plastic response without the TWIP effect. In Figure \ref{fig:fc-and-wq-modelling-combined}a, both modelled stress-strain and SHR curves are shown with the TWIP effect turned on or off. Remarkably, there was no significant loss in strain hardening even with the TWIP effect turned off. In the WQ condition (Figure \ref{fig:fc-and-wq-modelling-combined}b), there was almost no difference whether the TWIP effect was turned on or off. Therefore the modelled stress-strain and SHR curves with TWIP off were not shown in Figure \ref{fig:fc-and-wq-modelling-combined}b.

The component stress-strain curves in the FC and WQ conditions are shown in Figures \ref{fig:fc-and-wq-modelling-combined}e-f respectively. \textcolor{black}{In the FC condition,} the TWIP effect was observed to strengthen the austenite phase by 167 MPa at the failure strain. However, this strength was reduced to only 44 MPa \textcolor{black}{when the loss of austenite volume fraction to $\alpha'$-martensite transformation was taken into account} as seen in Figure \ref{fig:fc-and-wq-modelling-combined}g. \textcolor{black}{From Figure \ref{fig:fc-and-wq-modelling-combined}f, the austenite phase was not observed to strain harden significantly in the WQ condition, largely due to the close competiton between dislocation multiplication and annihilation arising from the short lamella thickness. The lack of any difference in strength between the TWIP on and off conditions in WQ was likely due to the severe reduction in the dislocation storage rate in the austenite grains due to the already very fine lamella widths \cite{Bouaziz2010}.} Therefore, further refinement of the grain size through the dynamic Hall-Petch effect \cite{DeCooman2018,Bouaziz2012} due to twinning proved to be ineffective in improving the strength of the austenite phase in the WQ condition. 

By multiplying the component strength by the respective volume fraction of each phase, the cumulative contribution to global strength from each phase with strain can be obtained and is shown in Figures \ref{fig:fc-and-wq-modelling-combined}g-h. It can be seen that the strength contribution from the TWIP effect is dwarfed by the TRIP effect. The \textcolor{black}{strength from the} $\alpha'$-martensite phase contributes to nearly 50\% of the UTS in the FC condition and more than 50\% of the UTS in the WQ condition.

\section{Discussion}

\subsection{$\alpha'$-martensite nucleation and growth}


In order to attain a sustained high SHR and large elongations in TRIP-assisted steels, it is necessary to continuously form fine $\alpha'$-martensite in a steady manner over a wide strain range \cite{Wang2016,Herrera2011}. A common strategy is to create a spread in austenite stability \textit{via} grain size distribution \cite{Wang2016}, inhomogeneous Mn composition \cite{Li2016b} or texture \cite{Xu2017}. This leads to a spectral TRIP \cite{Wang2016} or discontinous TRIP effect \cite{Li2016b,Xu2017,Wang2020} where transformation begins and ends with the least and the most stable austenite grains respectively.

In both FC and WQ conditions, the austenite phase was chemically very stable against $\alpha'$-martensitic transformation due to the high Mn and C content (Table \ref{tab:TEM-EDScomp}). Chatterjee \textit{et al.} \cite{Chatterjee2007} showed that with such high C content\textcolor{black}{s}, formation of Strain-Induced Martensite (SIM) would be highly improbable. However, an applied stress can provide an additional mechanical driving force, $\Delta G\textsubscript{mech}$, for Stress-Assisted Martensite (SAM) transformation \cite{Das2011,Das2015}. In medium Mn steels, it is known that strain and therefore stress localises at the interphase boundaries during deformation due to the strength mismatch between austenite, ferrite and $\alpha'$-martensite \cite{Lee2014a}. The high stress localisation was likely to have been able to provide a sufficiently high mechanical driving force for SAM to nucleate at austenite grain boundaries as observed in Figures \ref{fig:firstsaddletem}, \ref{fig:secondpeaktem} and \ref{fig:secondpeaktkd}. Grain boundary SAM nucleation was also observed by Yen \textit{et al.} \cite{Yen2015} in a medium Mn steel with a submicron grain size. The stress-assisted nature of $\alpha'$-martensite nucleation may explain why an incubation strain was observed (Figure \ref{fig:ebsd-micros}l) as it would be necessary to build up a critical local stress at the austenite grain boundaries. However, the stress field at the grain boundary would decay rapidly towards the interior of the parent austenite grain and the driving force for SAM transformation would similarly diminish. The highly local stress concentration may explain why the $\alpha'$-martensite grains in both FC and WQ were very small as $\alpha'$-martensite cannot grow past the stress field. With additional deformation, $\alpha'$-martensite grains with favourable orientations with stress will grow (areas 2 and 3 in Figure \ref{fig:secondpeaktkd}), whereas repeated nucleation of $\alpha'$-martensite nucleating on top of each other will occur (Figures \ref{fig:secondpeaktkd}b-c). The phenomenon of $\alpha'$-martensite only being able to nucleate and grow within local stress concentrations keeps the $\alpha'$-martensite grains small and greatly extends the strain regime where TRIP occurs.



\subsection{Effects of microstructure on TWIP and TRIP}

The effects of grain size on twinning and martensitic transformation in austenite are well studied. In TWIP steels, there is an impression that grain size reduction generally increases the twinning stress \cite{DeCooman2018}. However, in many studies, reducing the grain size to 1-5 \textmu m  does not appear to negatively affect tensile properties and elongation, although it is acknowledged that the twinning stress was increased \cite{Gutierrez-Urrutia2010,Rahman2015}. From Table \ref{tab:TEM-EDScomp}, the grain size, measured as the equivalent circle diameter, of FC and WQ conditions were not significantly different. However, from Figures \ref{fig:firstsaddletem} and \ref{fig:secondpeaktem} it was observed that extensive twinning occured in stage II and III for FC but mostly in stage III for WQ. In WQ, twins were observed to propagate across the width of the lamellar grains (Figures \ref{fig:secondpeaktem}e-h), implying that the lamellar width (approximately 300 nm) should be considered rather than the equivalent circle diameter.  Since the austenite lamellar width in the WQ condition was significantly finer than that of the equiaxed grain diameter in FC, the twinning stress in the WQ condition would be much higher and would explain why twinning in lamellar austenite grains was delayed to a later stage compared to equiaxed austenite grains. 

In TRIP-assisted steels, it is well known that a decreasing grain size has a strong mechanical stabilisation effect on austenite, inhibiting the formation of $\alpha'$-martensite \cite{Yang2009,Lee2013d,Sohn2014a}. Additionally, blocky or equiaxed austenite is generally less stable than film or lamellar austenite in medium Mn steels \cite{Chiang2015,He2020}. Therefore, it should follow that the WQ samples would form less $\alpha'$-martensite than the FC samples at a similar strain since the lamellar width of the WQ samples was much finer than the grain size in the FC samples. However, from Figure \ref{fig:ebsd-micros}l, there was more $\alpha'$-martensite in the WQ sample than in the FC sample at failure. 

In this alloy, the austenite phase of both FC and WQ conditions were chemically very stable, dominating the mechanical stability term due to grain size refinement in Equation \ref{eq:Ms}, as seen in Table \ref{tab:TEM-EDScomp}. The effect of relative grain size difference on austenite stability between the equiaxed grain diameter in FC samples and the lamella grain width in WQ samples were therefore not expected to be significant. However, $\alpha'$-martensite nucleation was shown to be restricted to the austenite grain boundaries where there is a local concentration of stress. In the WQ samples, the lamellar grain morphology has a larger grain boundary area to volume ratio and therefore able to provide a larger number of nucleation sites for $\alpha'$-martensite to form. Therefore, it was likely that $\alpha'$-martensite was able to nucleate more easily in the WQ samples, resulting in a higher $\alpha'$-martensite fraction at failure. However, it is acknowledged that a simple explanation cannot fully capture the complexity of $\alpha'$-martensitic transformation in WQ. In Figures \ref{fig:secondpeaktkd}d-f, the fine austenite lamella in Area 1 and a wide lamellar austenite grain in Area 4 had different surface area to volume ratios yet both remained largely untransformed, suggesting that other effects such as texture, Schmid factors and stress shielding were possibly also involved \cite{Xu2017,He2020,Chiang2015,Zhang2013}. Nevertheless, in this alloy with a very high austenite stability where the formation of $\alpha'$-martensite was nucleation-limited, grain boundary area to volume ratio would have certainly played a significant role among the other factors known to contribute to $\alpha'$-martensite transformation.

\subsection{Strain hardening behaviour and constitutive modelling}

Perhaps the most striking observation from the modified constitutive model in Figure \ref{fig:fc-and-wq-modelling-combined} was the lack of strengthening contribution from the TWIP effect, especially in the WQ condition. Through the original constitutive model developed for successive TWIP$+$TRIP medium Mn steel, Lee and De Cooman \cite{Lee2014} showed that the TWIP effect was less effective in small austenite grains and concluded that strengthening from the TRIP effect was more pronounced compared to the TWIP effect. In this study, a similar conclusion was reached for the FC steel with an equiaxed microstructure. However, this study extends the concept \textcolor{black}{to include} the simultaneous TWIP$+$TRIP mechanism and for lamellar microstructures which showed a nearly complete lack of contribution to strengthening from the TWIP effect. 

\textcolor{black}{Given the lack of strengthening from the TWIP effect,} it is therefore unsurprising that the observed differences in twinning kinetics in FC and WQ conditions did not result in a significant difference in the strain hardening behaviour. Instead, the strain hardening behaviour was dominated by the TRIP effect. Furthermore, since both FC and WQ conditions were found to share a similar $\alpha'$-martensite nucleation and growth mechanism, it is reasonable to expect the strain hardening profiles to be very similar.


Since the TWIP effect does little in terms of strength for medium Mn steels, it is probably better to not pursue the TWIP$+$TRIP effect in alloy design. In order to enable the TWIP$+$TRIP effect, a relatively high SFE and austenite stability is needed \cite{Kwok2022b}. This is enabled by a combination of either low Mn and high C (current alloy), or high Mn and low C. If the TWIP effect can be forgone during alloy design, it would be possible to enable low to medium Mn and low C (3-6 wt\% Mn, 0.05-0.2 wt\% C) compositions that exhibit the TRIP effect only. \textcolor{black}{Such low Mn, low C compositions have been increasingly termed lean medium Mn steels \cite{Kaar2020b}, and are} desirable in terms of lower segregation after casting \cite{Kwok2022a}, better weldability \cite{Qi2018}, processability, \textit{etc}. However, it is worth noting that while the TWIP effect does little for TWIP$+$TRIP-type medium Mn steels, the tensile properties still tend to be better in terms of elongation than in pure TRIP-type medium Mn steels \cite{Lee2014,Sohn2014a,Sohn2017a,Xu2022}. The higher Mn and C contents necessary for the TWIP$+$TRIP effect also stabilises the austenite phase and prolongs the TRIP effect to significantly higher strains. Therefore, the alloy chemistry surrounding TWIP$+$TRIP-type medium Mn steels \textcolor{black}{is} still worth further study, although less focus might be given to the TWIP effect. \textcolor{black}{It may also be of research interest to determine if the TWIP effect may have other benefits in terms of ductility or performance at high strain rates.}

\section{Conclusion}

The effects of microstructure on the TWIP$+$TRIP mechanism were examined in a 5Mn-0.5C type medium Mn steel in two conditions representing different processing strategies in the steel mill. In the first \textcolor{black}{condition}, furnace cooling after hot rolling was employed (FC), as in a coiler; and in the other, water quenching (WQ) on the run-out bed. Both were then intercritically annealed, as on a continuous annealing line. These \textcolor{black}{processing strategies} produced steels with similar initial austenite compositions and phase fractions but with equiaxed vs mixed equiaxed$+$lamellar microstructures, respectively. \textcolor{black}{The main findings are:}

\begin{enumerate}
	
\item Both FC and WQ conditions showed superior mechanical properties \textcolor{black}{compared to TWIP and DP steels}, and the simultaneous TWIP$+$TRIP plasticity enhancing mechanism was identified to be operative in both conditions regardless of microstructural form.

\item A novel $\alpha'$-martensite nucleation and growth mechanism in high C austenite was proposed. Stress-assisted $\alpha'$-martensite was able to nucleate at the austenite grain boundaries due to the high stress concentration during deformation. The $\alpha'$-martensite is unable to grow beyond the stress field into the parent austenite grain due to the high chemical stability and therefore remains small. Therefore, the continuous formation and slow growth of $\alpha'$-martensite grains greatly extends the strain regime where TRIP is operative and allows for large elongations to failure.

\item The shorter austenite lamella width in the WQ samples compared to the austenite grain diameter in the FC samples resulted in a shorter mean free path for twinning. This raised the critical twinning stress such that extensive twinning was only observed at higher strains in the WQ sample. 

\item A modified constitutive model was developed and found to have a good fit with the experimental data. The model also showed that the TWIP effect did not provide significant strengthening in the FC condition and almost no strengthening in the WQ condition. \textcolor{black}{The lack of strengthening from the TWIP effect was attributed to the extremely fine slip length in the austenite phase, especially in WQ condition where the lamella thickness was in the submicron regime.}


\end{enumerate}

\subsection{Acknowledgements}

TWJK would like to thank A*STAR, Singapore for a studentship. PG would like to acknowledge funding from SUSTAIN Future Steel Manufacturing Research Hub (EP/S018107/1). DD's work in the early stages of this project was funded by EPSRC (EP/L025213/1) and he also holds a Royal Society Industry Fellowship.

\section{Appendix}

To model the simultaneous TWIP$+$TRIP effect in the FC and WQ conditions, we follow the constitutive modelling work of Lee \textit{et al.} \cite{Lee2014,Lee2015b,Lee2013b} and Latypov \textit{et al.} \cite{Latypov2016}. This work similarly applies the iso-work assumption to model the strain partitioning between austenite, ferrite and martensite. The iso-work assumption can be expressed as:

\begin{equation}
\sigma_\gamma d \varepsilon_\gamma = \sigma_\alpha d \varepsilon_\alpha = \sigma_{\alpha'} d \varepsilon_{\alpha'}
\end{equation}

where $\sigma_\gamma$, $\sigma_\alpha$ and $\sigma_{\alpha'}$ are the flow stresses of austenite, ferrite and $\alpha'$-martensite respectively and $d \varepsilon_\gamma$, $d \varepsilon_\alpha$ and $d \varepsilon_{\alpha'}$ are the incremental strains of austenite, ferrite and $\alpha'$-martensite respectively. The global applied strain can therefore be expressed as a rule of mixtures:

\begin{equation}
d\varepsilon = f_\gamma d \varepsilon_\gamma + f_\alpha d \varepsilon_\alpha + f_{\alpha'} d \varepsilon_{\alpha'}
\end{equation}

where $f_\gamma$, $f_\alpha$ and $f_{\alpha'}$ are the phase fractions of austenite, ferrite and $\alpha'$-martensite respectively. Stresses were calculated by incrementally increasing true strain and updating the dislocation densities using the local gradient of dislocation density as a function of strain. To calculate strain partitioning, iso-work constraints were applied by using a trust-region-dogleg algorithm, a refined Newton's method in MATLAB. The model was implemented in MATLAB and is available online \cite{Rose2022}.

To begin, the flow stress, $\sigma$, of FC and WQ was assumed to obey the law of mixtures:

\begin{equation}
\sigma = \sigma_\gamma f_\gamma + \sigma_\alpha f_\alpha + \sigma_{\alpha'} f_{\alpha'}
\end{equation}

The flow stress of each phase, $\sigma_i$, where $i$ denotes austenite, ferrite or $\alpha'$-martensite, can be described as:

\begin{equation}
\sigma_i = \sigma_i ^{YS} + A M_i \mu_i b_i \sqrt{\rho_i}
\end{equation}

where  $\sigma_i^{YS}$ is the yield strength, $A$ is a constant equal to 0.4, $M_i$ is the Taylor factor, $\mu_i$ is the shear modulus, $b_i$ is the magnitude of the Burger's vector and $\rho_i$ is the stored dislocation density of phase $i$. The yield strength of each phase was determined by summing the solid solution and grain size strengthening contributions through the following equation:

\begin{equation}
\sigma_{i}^{YS} = \sigma^s_{i} + \frac{K_{i}}{\sqrt{d_{i}}}
\end{equation}

where $\sigma_i^s$ is the solid solution strength, $K_i$ is the Hall-Petch parameter and $d_i$ is the grain size of phase $i$. The austenite Hall-Petch parameter by Rahman \textit{et al.} \cite{Rahman2015} was used in place of original value used by Lee and De Cooman \cite{Lee2015b} as it was found to give a better fit. Additionally, the austenite lath width in the WQ sample was used for $d_\gamma$ rather than the ECD grain size in Table \ref{tab:TEM-EDScomp}. The solid solution strength of each phase was calculated according to the following empirical equations \cite{DeCooman2011a,Speich1968}:

\begin{equation}
\sigma^s_{\gamma} = 567(X^{\gamma}_C) - 1.5(X^{\gamma}_{Mn}) + 23 (X^{\gamma}_{Si})
\end{equation}

\begin{equation}
\sigma^s_{\alpha} = 5000 (X^{\alpha}_C) + 44.7 (X^{\alpha}_{Mn}) + 138.6 (X^{\alpha}_{Si})+ 70 (X^{\alpha}_{Al})
\end{equation}

\begin{equation}
\sigma^s_{\alpha'} = 413 + 1720 (X^\gamma_C) 
\end{equation}

where $X_j^i$ is the concentration of element $j$, in mass percent, in phase $i$. The evolution of dislocation density with strain, $\frac{d\rho_{i}}{d\epsilon_{i}}$, was determined by calculating the rate of dislocation storage and annihilation in each phase using the Kocks-Mecking model given as \footnote{It should be noted that Equation \ref{eq:disl_evol} contained a typographical error in the original reference \cite{Lee2014}. We have verified this with the authors of that paper and rectified this in the current paper.}:

\begin{equation}
\label{eq:disl_evol}
\frac{d\rho_{i}}{d\varepsilon_{i}} = M \left( \frac{P_{i}}{b_{i}\Lambda_{i}} + \frac{k^{i}_1}{b_{i}} \sqrt{\rho_{i}} - k^{i}_2 \rho_{i} \right)
\end{equation}

where $P_i$ is a coefficient related to the grain size, $\Lambda_i$ is the dislocation mean free path, $k_1^i$ is the dislocation storage coefficient and $k_2^i$ is the dislocation annihilation coefficient of phase $i$. Here, we \textcolor{black}{approximate the strength of $\alpha'$-martensite to remain constant with strain based on the findings by Latypov \textit{et al.} \cite{Latypov2016} and} set $k_1^{\alpha'}$ and $k_2^{\alpha'}$ to zero. The term $P_i$ is defined as the probability for a dislocation to not be absorbed into a grain boundary \cite{Bouaziz2010} and is given as:

\begin{equation}
P_{i} = exp \left[- \left(\frac{d^c_{i}}{d_{i}}\right)^3 \right]
\end{equation}

\begin{table}[t]
	\centering
	\caption{Table of parameter values used in the constitutive model after the work by Lee and De Cooman \cite{Lee2014}.}
	\begin{adjustbox}{width=\columnwidth,center}
		\begin{tabular}{lcccc}
			\toprule
			& Units & Austenite & Ferrite & Martensite \\
			\midrule
			$\mu_i$ & GPa    & 75  & 75 & 80  \\
			$K_i$    & MPa \textmu m$^{-0.5}$ & 330 \cite{Rahman2015} & 172 \cite{Harding1969} & 0 \\
			$b_i$    & $\angstrom$     & 2.50  & 2.48 & 2.48 \\
			$k_1$    & - & 0.006 \cite{Latypov2016} & 0.0065 \cite{Latypov2016}  & 0  \\
			$k_2$    & -     & 2 \cite{Latypov2016} & 2 \cite{Latypov2016} & 0 \\
			$d_i^c$   & \textmu m     & 1.5 \cite{Lee2013b} & 2.1 \cite{Lee2013b} & 2.1 \cite{Lee2013b} \\
			$\rho_{ini}$ & m$^{-2}$ & $10^{12}$ & $10^{12}$ & $10^{15}$ \\
			$\alpha$   &   -    &   3.5    &  -     & - \\
			$m$     &   -    & 2     &   -    & - \\
			$f_0$    &   -    & 0.2 \cite{Bouaziz2008}  &   -    & - \\
			$M$     & - & 3.06  & 2.95  & 3.06 \\
			\bottomrule
		\end{tabular}%
	\end{adjustbox}
	\label{tab:listofconstants}%
\end{table}%

where $d_c^i$ is a critical grain size of phase $i$, below which the rate of dislocation annihilation at the grain boundaries is larger than the rate of dislocation storage and \textit{vice versa}. The dislocation mean free path of ferrite, $\Lambda_\alpha$, and $\alpha'$-martensite, $\Lambda_{\alpha'}$, was assumed to be equal to the average grain size. However, for the austenite phase, the dislocation mean free path, $\Lambda_\gamma$ was determined as:

\begin{equation}
\Lambda_{\gamma} = \left( \frac{1}{d_{\gamma}} + \frac{1}{\lambda_{T}} \right)^{-1}
\end{equation}

where $\lambda_{T}$ is the mean twin spacing and is given as the following equation:

\begin{equation}
\lambda_T = 2c_T \frac{1-f_T}{f_T}
\end{equation}

where $c_T$ is the twin thickness and $f_T$ is the volume fraction of twins. The evolution of twin volume fraction with strain can be expressed as:

\begin{equation}
f_T = f_\gamma f_0 [1-\exp(-\alpha \varepsilon)]^m
\end{equation}

where $f_0$ is the twin saturation volume fraction, $\alpha$ is the coefficient associated with the formation of a twin nucleus when perfect dislocations intersect and $m$ is the exponent associated with the probability of perfect dislocations intersecting. Finally, since the evolution of $\alpha'$ martensite with strain no longer relies on the prior formation of twins, a simple Avrami equation \cite{Avrami1939, Avrami1940, Avrami1941} was used to describe the evolution of martensite fraction with strain:

\begin{equation}
\label{eq:Avrami}
f_{\alpha'} = (1-f_\alpha) (1-\exp(-a \varepsilon^b))
\end{equation}

where $a$ and $b$ are Avrami constants. Since precise measurement of martensite fraction was not possible, the Avrami constants were fitted by minimising the \textchi$^2$ of the modelled and experimental strain hardening data using a gradient search. Values for the parameters used for the model are given in Table \ref{tab:listofconstants} and \ref{tab:listofconstants_exp}.

\begin{table}[t]
	\centering
	\caption{Table of experimental and fitting parameters used in the constitutive model.}
	\begin{adjustbox}{width=\columnwidth,center}
		\begin{tabular}{lcccc}
			\toprule
			& Units & Austenite & Ferrite & Martensite \\
			\midrule
			$c_T$  & nm     & 30 & - & - \\
			$d_i$  (FC)  & \textmu m     & 1.5 & 3.0 & 0.2 \\
			$d_i$ (WQ) & \textmu m     & 0.3 & 1.3 & 0.2  \\
			$a$ (FC)&		-					&	-	&	-	& 3.3 \\
			$a$ (WQ)&		-					&	-	&	-	& 3.7 \\
			$b$ (FC)&		-					&	-	&	-	& 3.1 \\
			$b$ (WQ)&		-					&	-	&	-	& 2.5 \\			
			\bottomrule
		\end{tabular}%
	\end{adjustbox}
	\label{tab:listofconstants_exp}%
\end{table}%

\newpage



\end{document}